\documentclass[a4paper,11pt]{article}
\pdfoutput=1
\usepackage{jcappub}
\usepackage[english]{babel}
\usepackage[utf8]{inputenc}

\usepackage{multirow}
\usepackage{textcomp,color}
\usepackage{graphicx,subfigure}

\usepackage[normalem]{ulem}

\DeclareMathOperator{\diag}{diag}

\title{Classical transitions with the topological number changing in the early Universe}

\author[a,b,1]{Vakhid A. Gani,\note{Corresponding author.}}
\author[a,c]{Alexander A. Kirillov,}
\author[a,d]{Sergey G. Rubin}

\affiliation[a]{National Research Nuclear University MEPhI\\ (Moscow Engineering Physics Institute), 115409 Moscow, Russia}
\affiliation[b]{Theory Department, National Research Center Kurchatov Institute,\\ Institute for Theoretical and Experimental Physics, 117218 Moscow, Russia}
\affiliation[c]{Department of Theoretical Physics, Yaroslavl State P.G. Demidov University,\\ 150003 Yaroslavl, Russia} 
\affiliation[d]{N.I. Lobachevsky Institute of Mathematics and Mechanics, Kazan Federal University,\\ Kremlevskaya  street 18, 420008 Kazan, Russia}

\emailAdd{vagani@mephi.ru}
\emailAdd{aakirillov@mephi.ru}
\emailAdd{sergeirubin@list.ru}

\abstract{We consider classical dynamics of two real scalar fields within a model with the potential having a saddle point. The solitons of such model are field configurations that have the form of closed loops in the field space. We study the formation and evolution of these solitons, in particular, the conditions at which they could be formed even when the model potential has only one minimum. These non-trivial field configurations represent domain walls in the three-dimensional physical space. The set of these configurations can be split into disjoint equivalence classes. We provide a simple expression for the winding number of an arbitrary closed loop in the field space and discuss the transitions that change the winding number. We also show that non-trivial field configurations could be responsible for the energy density excess that could evade the CMB constraints but could be important at scales  which are responsible for the formation of galaxies and the massive primordial black holes.}

\makeatletter
\def\@fpheader{\relax}
\makeatother

\begin{document}

\maketitle

\flushbottom

\section{Introduction}
\label{sec:Introduction}

Inflationary models with several fields have recently become the subject of a growing interest \cite{Dias,Bramante,Abbyazov}. Most of them are based on complicated potentials, which arise in the brane approach \cite{Liddle0, Lust}, various supersymmetric models \cite{Kawai}, and the landscape paradigm, see, e.g., \cite{susskind01,Vil,vilenkin.jpa.2007,denef.annphys.2007,douglas.jhep.2003}.

Multi-field inflation has been thoroughly elaborated, with its influence on CMB fluctuations analyzed in, e.g., \cite{Liddle,Bach,li01,duplessis01}. An important feature of multi-field inflationary models is a potential with many minima \cite{Liu.2015}. It is well-known that a field-theoretical system with such a potential can admit non-trivial configurations of the scalar field --- solitons \cite{manton}. After the end of the inflation such configurations could collapse into black holes \cite{Konopl, RubKhlopSakh_PBHProd, KhlopRubSakh} or form domains with extra energy density. Remarkably, solitons could be formed even if the model potential has only one minimum \cite{manton}. This work considers the conditions at which this can occur.

As shown in \cite{Aazami}, the number of saddle points of the multi-field potential can be much greater than the number of its minima. Therefore, if the horizon of the Universe was formed near an appropriate minimum of the potential \cite{loeb.jcap.2006,garriga.ptps.2006}, there is a substantial probability to have its saddle point(s) which has to be taken into account. This provides additional motivation to study the saddle point influence on the inflationary dynamics. Of the plethora of the inflationary models, those with potentials having saddle points have important implications. They include the models of hybrid inflation \cite{Linde_Hyb_Infl} and their modifications, and other modern models. For example, the Aligned Natural Inflation was recently shown to have higher altitude inflationary trajectories passing through the saddle points of the two-field potential \cite{Peloso.JCAP.2015}; the influence of a nearby saddle point on the inflationary process was also studied in field spaces of higher dimensions \cite{Battefeld.JCAP.2013}.

Our paper considers a system with two real scalar fields, denoted as $\varphi$ and $\chi$. We show that the domain wall formation is possible even if the potential has only one minimum. We demonstrate that non-trivial field configurations can occur when the fields reach the same minimum at different spatial asymptotics. The necessary condition for the realization of this phenomenon is the existence of a saddle point of the potential. We perform the topological classification of such configurations, which allows us to divide them into disjoint homotopic classes. We also investigate the conditions at which the configurations can move from one homotopic class to another. Note that such transitions are forbidden if the fields reach different minima of the potential at different spatial asymptotics; in this case only quantum transitions due to tunneling are allowed \cite{Dine}.

Our paper is organized as follows. In section \ref{sec:Qualitative} we give some qualitative analysis of how the initial configuration could be formed. In section \ref{sec:Configurations} we describe the model and the approximations employed by us, and introduce the homotopic classification of the field configurations. In section \ref{sec:Numerical} we show the results of our numerical simulation of the evolution of selected initial configurations and discuss the change of the winding number during the classical field evolution. Section \ref{sec:Models} presents a discussion of the primordial black holes (PBHs) formation within the modern inflationary models. In section \ref{sec:Conclusion} we conclude with a brief discussion of the results and the prospects for future work.

\section{Qualitative analysis}
\label{sec:Qualitative}

The arguments regarding the field dynamics and formation of the initial conditions can be set out as follows. During the inflation, the fields $\varphi$ and $\chi$ undergo both quantum and classical evolution. At this stage, the quantum fluctuations of $\varphi$ and $\chi$ are of the order of $\delta\varphi\simeq\delta\chi\simeq H_\text{I}/2\pi$, where $H_\text{I}$ is the Hubble parameter during the inflation \cite{Star}. These fluctuations lead to a highly inhomogeneous fields distribution soon after the beginning of the inflation. This leads to the large scale structure formation after the end of the inflation if the inflaton potential has an ordinary form like quadratic one or similar to it.  The picture changes drastically if the potential is assumed to be more complex. As was mentioned in the Introduction, the potentials with several minima lead to the primordial black holes (PBHs) formation that is topical nowadays. In this section we qualitatively show that the potentials with one minimum and saddle point(s) can also lead to the PBH formation after the end of the inflation.

Let us suppose that the space under the modern horizon is filled by the field value near a saddle point of the potential. The space is split up into many causally disconnected regions, the number of which depends on a specific e-fold. Due to quantum fluctuations, the fields in one of these regions, $\mathcal B$, could overcome the saddle point, while in the other part of the space, $\mathcal U$, the fields remain on the same side of the saddle point. Later on the fields in the region $\mathcal B$ roll down to the potential minimum. The region $\mathcal B$ is surrounded by the region $\mathcal U$, where the fields are on the other side of the saddle point. Let this fluctuation happens at the e-fold number $N$. Then starting from the e-fold number $N+1$  internal points of the domains  $\mathcal B$  and  $\mathcal U$ are causally disconnected because the distance between them becomes larger than the horizon $H_\mathrm{I}$. This means that the fields inside these domains evolve independently.

The potential with a saddle point determines the steepest descent line $l_\text{sd}$ passing through the saddle point $(\varphi_\text{s},\chi_\text{s})$. The space region $\mathcal B$ contains the fields on the one side of the saddle point while the fields in another region $\mathcal U$ are placed on another side of the saddle point. The space boundary $\partial\mathcal B$ of the domain $\mathcal B$ consists of points where $\varphi = \varphi_\text{s}$, $\chi =\chi_\text{s}$, and therefore the potential is maximal at those points  of the line $l_\text{sd}$, which belong to the border $\boldsymbol{\partial}\mathcal{B}$. 

In the process of further evolution the fields in $\mathcal B$ and $\mathcal U$ will roll down into the potential minimum $(\varphi_\text{min},\chi_\text{min})$, but their trajectories will lie on the different sides of the saddle point. As a result, after a long period of independent evolution the fields arrive to a configuration that encircles a local maximum of the potential. The values of the fields at spatial points $x$ far from the boundary $\partial\mathcal B$ are close to $(\varphi_\text{min},\chi_\text{min})$. The line $l_\text{sd}$ contains the saddle point $(\varphi_\text{s},\chi_\text{s})$ with both its ends arriving at the same minimum $(\varphi_\text{min},\chi_\text{min})$. Therefore, starting at one end of $l_\text{sd}$ and moving to the other, the fields' values have to pass through $(\varphi_\text{s},\chi_\text{s})$ that defines the border $\partial\mathcal B$, which thus becomes a domain wall. The horizon grows quickly after the inflation is finished. When the horizon crosses size of the domain wall, the latter start shrinking finally producing PBHs. This phenomenon was discussed in ~\cite{RubKhlopSakh_PBHProd,KhlopRubSakh,Dokuch_Quasars} for potentials with several minima.
 
In this paper we discuss a set of potentials with saddle points and one minimum. In this case the domain walls not only shrink but their energy density is fluctuating and producing coherent field oscillations. Our numerical simulations, discussed in detail in what follows, confirm this qualitative picture and reveal the conditions at which these arguments hold.

Initial setup considers the fields evolving in the $(3+1)$-dimensional space-time. Here we will study a particular solutions in the form of the one-dimensional kinks. Multi-field inflationary models could thus potentially benefit from connections to $(1+1)$-dimensional field-theoretical models with potentials with one or more minima, where many interesting and important results have been obtained recently, in particular, related to interactions of solitary waves \cite{GaKuPRE,GaKuLi,Aliakbar,Demirkaya,christov01,GaLeLi,saadatmand01,GaLeLiconf,simas01,weigel01,Weigel.conf.2017,Weigel.PLB.2017,Belendryasova.arXiv.08.2017,Belendryasova.conf.2017,Bazeia.arXiv.2017.sinh,Bazeia.arXiv.2017.sinh.conf,Gani.arXiv.2017.dsg,ashcroft01,Radomskiy}, solitons and domains stability \cite{kurochkin,godunov,godunov02,Gani.YaF.1999,Gani.YaF.2001}, and planar domain walls \cite{GaKu.SuSy.2001,GaKsKu01,GaKsKu02,lensky,brihaye01,GaLiRa,GaLiRaconf}, see also review \cite{aek01} and books \cite{manton,vilenkin01}.

\section{Nontrivial field configurations and their classification}
\label{sec:Configurations}

\subsection{The model}

We consider two models with potentials characterized by one minimum and one saddle point. It is assumed that classical trajectories pass near a maximum of the potential and its saddle point even in multifield inflation. One of the potentials, eq.~\eqref{eq:potential}, has quite general form with several parameters. The other one, eq.~\eqref{eq:Potential_Mex}, represents well known form of the tilted Mexican hat. Both of them are endowed with a saddle point.

Assume that the inflation is finished and the horizon crosses the size of the domain $\mathcal{B}$. Consider a model with two real scalar fields $\varphi$ and $\chi$ in the (3+1)-dimensional space-time. The dynamics of the field system is determined by the Lagrangian
\begin{equation}
	\label{eq:Lagrangian}
	\mathcal{L} = \frac{1}{2} g^{\mu\nu} \left(  \partial_{\mu}\varphi\partial_{\nu}\varphi + \partial_{\mu}\chi\partial_{\nu}\chi \right)
    	- V(\varphi,\chi),
\end{equation}
where the metric tensor $g_{\mu\nu}$ for the Friedmann-Robertson-Walker universe is
\begin{gather}
	\label{eq:g}
	g_{\mu\nu} = \diag\left( 1, -a^2(t), -a^2(t)\,r^2 , -a^2(t)\, r^2\sin^2\theta \right).
\end{gather}
The equations of motion following from the Lagrangian \eqref{eq:Lagrangian} are
\begin{gather}
	\label{eq:MotEqs}
	 \Box \varphi = -\frac{\partial V}{\partial\varphi},\quad 
     \Box \chi = -\frac{\partial V}{\partial\chi},
\end{gather}
where $\Box = \displaystyle\frac{1}{\sqrt{-g}}\, \partial_\mu\left(\sqrt{-g}\; g^{\mu\nu} \partial_\nu\right)$ is the d'Alembert operator. Taking into account the metric \eqref{eq:g}, we obtain:
\begin{equation}
	\begin{cases}
		\varphi_{tt} + 3 H \varphi_t - a^{-2}(t)\,\varphi_{rr} 
        	-\cfrac{2}{r}\, a^{-2}(t)\, \varphi_r 
            = -\cfrac{\partial V}{\partial\varphi},\\
		\chi_{tt} + 3 H \chi_t - a^{-2}(t)\,\chi_{rr} 
        	-\cfrac{2}{r}\, a^{-2}(t)\, \chi_r 
            = - \cfrac{\partial V}{\partial\chi},
	\end{cases}
    \label{eq:Equations_1}
\end{equation}
where the Hubble parameter $H=\dot{a}/a$ is small after the end of the inflation. The friction term is not necessary, however, it provides the dumping of long-period oscillations during the reheating stage. We are looking for a domain $\mathcal B$, which is surrounded by the border $\partial\mathcal B$ with the fields' values at the saddle point $(\varphi_\text{s},\chi_\text{s})$. As we show below, a nontrivial field configuration can be formed even if the vacuum state is the same at both sides of the border.

During the inflation, the size of the domain $\mathcal B$ significantly increases. Hence, for an observer near the border $\partial\mathcal{B}$ the latter is almost flat, and one can restrict the problem to the case of flat solutions $\varphi(x,t), \chi(x,t)$ depending on one spatial coordinate $x$. The equations of motion \eqref{eq:Equations_1} can also be transformed in this case in such a way that they only depend on $t$ and $x$. The terms $\displaystyle\frac{2}{r}a^{-2}(t)\,\varphi_r$ and $\cfrac{2}{r}a^{-2}(t)\,\chi_r$ can also be omitted as it is usually done within the thin wall approximation. In the following we use the physical distances $R=a(t)r$ and assume that the cosmological expansion is small. In this approximation equations \eqref{eq:Equations_1} become
\begin{equation}
	\begin{cases}
		\varphi_{tt} + 3 H \varphi_t - \varphi_{xx} = - \displaystyle\frac{\partial V}{\partial\varphi},\\
		\chi_{tt} + 3 H \chi_t - \chi_{xx} = - \displaystyle\frac{\partial V}{\partial\chi}.
	\end{cases}
    \label{eq:Equations}
\end{equation}

\begin{figure}[t]
 \centering
 \subfigure{
  \includegraphics[width=0.47\textwidth]{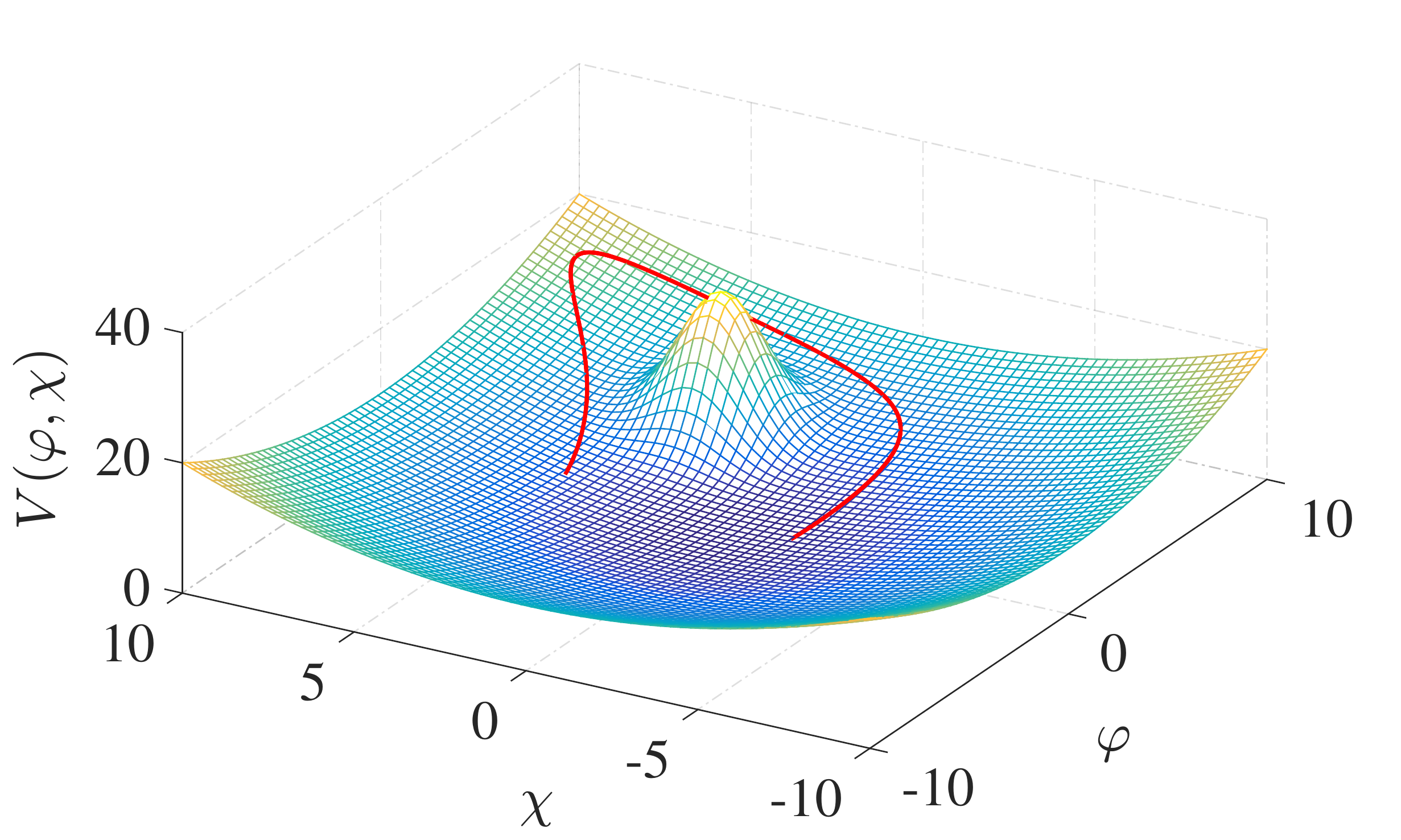}
 }
 \subfigure{
  \includegraphics[width=0.47\textwidth]{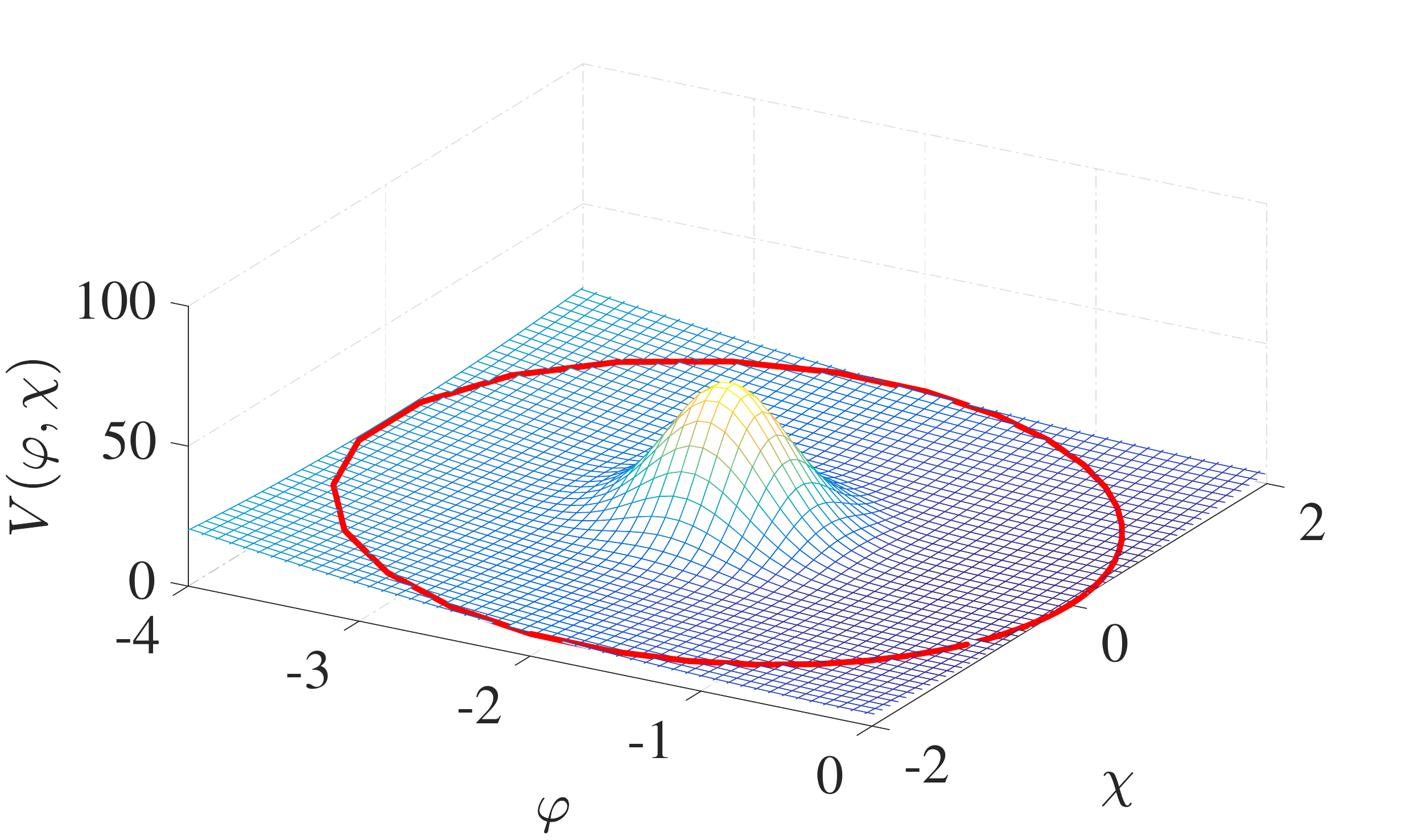}
 }
 \caption{Left panel: one of the possible initial field configurations. Right panel: initial field configuration \eqref{eq:InCond} which is used in our calculations. All numbers are in the $H_\mathrm{I}$ units.
 }
\label{fig:Potential+InCond}
\end{figure}

We take the potential $V(\varphi,\chi)$ with a saddle point in the form
\begin{equation}
	\label{eq:potential}
	V(\varphi,\chi) = d\,(\varphi^2 + \chi^2) + 
    	a\:\exp{[-b\:(\varphi-\varphi_0^{})^2 - c\:(\chi-\chi_0^{})^2]},
\end{equation}
where $a>0$, $b>0$, $c>0$, $d>0$ and $\varphi_0^{}$, $\chi_0^{}$ are the physical parameters of the model. The parameters $\varphi_0^{}$ and $\chi_0^{}$ fix the position of the maximum of the potential, while $a$ defines the height of the maximum. The constants $b$ and $c$ describe the shape of the maximum. The typical shape of the potential \eqref{eq:potential} is shown in figure \ref{fig:Potential+InCond}.
With exponentially small errors, the minimum of the potential \eqref{eq:potential} is at the point $(0,0)$. The value of $H_\mathrm{I}\sim10^{13}\,\text{GeV} \sim 10^{-6} M_\mathrm{Pl}$ gives the natural scale of the initial field configurations. Below we use $H_\mathrm{I}$ units for numerical simulation.

\subsection{The winding number}

Field configuration at any moment of time is a pair of smooth functions $\varphi(x)$, $\chi(x)$, which map the physical space $x$ to the field space $(\varphi,\chi)$. We can imagine that the point $(\varphi,\chi)$ is on the potential surface \eqref{eq:potential}. Thus the field configuration at any given time moment can be visualized as a curve on the potential surface \eqref{eq:potential}.

As it has already been mentioned, the initial field configurations are the results of quantum fluctuations during the inflation. In the post-inflationary period, the initial values of the fields in causally disconnected domains substantially differ. After the inflation the horizon starts to grow, and many such domains become causally connected. The initial field configuration in any space domain is unique and can correspond to an arbitrary smooth curve in the $(\varphi,\chi)$ plane, see, e.g., figure \ref{fig:Potential+InCond} (left panel). The initial curve chosen there as an example is not closed. Nevertheless, both ends of this curve tend to evolve to the same point~--- the potential minimum, resulting in a final configuration represented by a closed curve.

The set of all closed curves in $(\varphi,\chi)$ plane can be split into homotopic classes (equivalence classes). The equivalent curves are those having the same winding number $N$ --- the number of turns of the curve around the point $(\varphi_0^{},\chi_0^{})$. An explicit formula for $N$ can be obtained using the residue theorem: the line integral of a function $f(\zeta)$ of the complex variable $\zeta=\varphi+i\chi$ around the curve is equal to $2\pi i$ times the sum of residues of $f(\zeta)$ at isolated singular points, each counted as many times as the curve winds around the point. We use the following simple function:
\begin{equation}
f(\zeta)=\frac{1}{\zeta-\zeta_0^{}},
\end{equation}
which has a simple pole at $\zeta_0^{}=\varphi_0^{}+i\chi_0^{}$, and the residue of $f$ at this point is equal to $1$. This results in
\begin{equation}
N = \frac{1}{2\pi i}
\oint f(\zeta)d\zeta = \frac{1}{2\pi i} \oint \frac{d\zeta}{\zeta-\zeta_0^{}},
\end{equation}
where the contour integration is performed along the closed curve, which winds $N$ times around the pole. Substituting $\zeta=\varphi+i\chi$, $\zeta_0^{}=\varphi_0^{}+i\chi_0^{}$ and simplifying the expression, we obtain
\begin{equation}
	N[\varphi,\chi] = \frac{1}{2\pi}\int\limits_{-\infty}^{+\infty}
    \frac{(\varphi-\varphi_0^{})
	\chi_x -(\chi-\chi_0^{})
	\varphi_x}{(\varphi-\varphi_0^{})^2+(\chi-\chi_0^{})^2}\:dx.
    \label{eq:NumWound}
\end{equation}
We have thus separated the set of all closed curves in the $(\varphi,\chi)$ plane into homotopic classes, depending on the number of windings of the curve around the potential maximum $(\varphi_0^{},\chi_0^{})$. Depending on the shape of the potential, one of the following ways could be realized:
\begin{enumerate}
\item The height of the maximum of the potential at the point $(\varphi_0^{},\chi_0^{})$ is negligible or absent, and any trajectory contracts to the point $(\varphi_\text{min},\chi_\text{min})$ as the result of the classical evolution.
The evolution of any configuration thus turns it into the trivial configuration, i.e.\ $\varphi(x,t\to+\infty)\to\varphi_\text{min}$, $\chi(x,t\to+\infty)\to\chi_\text{min}$. In this case the splitting into the homotopic classes does not make sense.
\item The height of the maximum of the potential at the point $(\varphi_0^{},\chi_0^{})$ is infinite, and the winding number $N$ is conserved during the classical dynamics of any closed trajectory.
\item The maximum of the potential has a moderate height. As we demonstrate below, in this case the winding number of any configuration can be conserved or can decrease, depending on the initial conditions and the parameters of the model.
\end{enumerate}

\section{Numerical simulation. Formation of the domain wall(s)}
\label{sec:Numerical}

The above qualitative analysis shows that the presence of saddle point(s) could lead to the formation of the non-trivial field configurations. In this section, we perform numerical simulation to validate this analysis. We will consider the third scenario from the list above. Studying the classical evolution of the fields, we show that their dynamics can vary substantially, depending on the initial configuration and the values of the model parameters.

We start from an initial field configuration in the form of a closed loop in the $(\varphi,\chi)$ plane, so that the boundary conditions are the following:
\begin{equation}
	\begin{cases}
		\varphi(-\infty,t) = \varphi(+\infty,t),\\
        \varphi_x(-\infty,t) = \varphi_x(+\infty,t),\\
		\chi(-\infty,t) = \chi(+\infty,t),\\
        \chi_x(-\infty,t) = \chi_x(+\infty,t).
	\end{cases}
    \label{eq:BondCond}
\end{equation}
Furthermore, we specifically consider the following initial configuration which encircles the local maximum of the potential \eqref{eq:potential}:
\begin{equation}
	\begin{cases}
		\varphi(x) = \varphi_0^{} + R\cos\theta(x),\\
		\chi(x) = \chi_0^{} + R\sin\theta(x),
	\end{cases}
    \label{eq:InCond}
\end{equation}
where the dependence $\theta(x)$ is given by
\begin{equation}
	\theta(x) = N \pi \left( 1 + \tanh\left(\frac{x}{l}\right)\right).
    \label{eq:Theta}
\end{equation}
Our initial trajectory is parameterized by $R$ and $l$. In our estimates we put $l=1$ (remind that $H_\mathrm{I}=1$).

It is easy to see that $\theta$ varies within the interval $0\le\theta(x)\le 2\pi N$ at $-\infty\le x\le+\infty$. Note also that this initial trajectory fulfills the boundary conditions \eqref{eq:BondCond}. We show this trajectory in figure \ref{fig:Potential+InCond} (right panel). Substituting \eqref{eq:InCond} and \eqref{eq:Theta} in \eqref{eq:NumWound} gives the correct answer for the winding number, equal to the coefficient $N$ in \eqref{eq:Theta}.

\begin{figure}[th!]
\begin{center}
  \subfigure[The case of $N=1$.]{
  \includegraphics[width=0.47\textwidth]{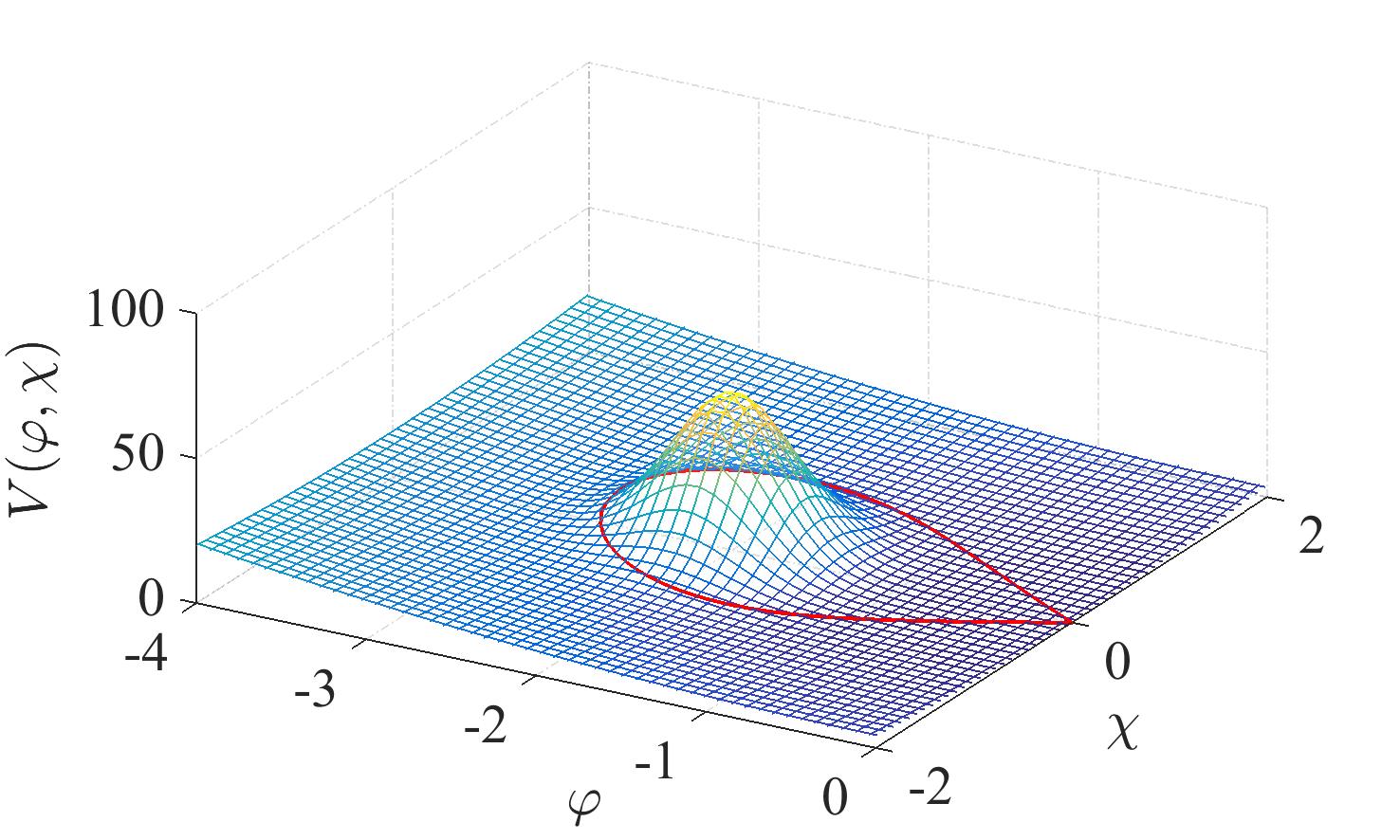}
	  \label{fig:Potential-1N}
	 }
  \subfigure[The case of $N=2$. For $N>2$ the number of turns of the curve around the local maximum is equal to $N$.]{
	  \includegraphics[width=0.47\textwidth]{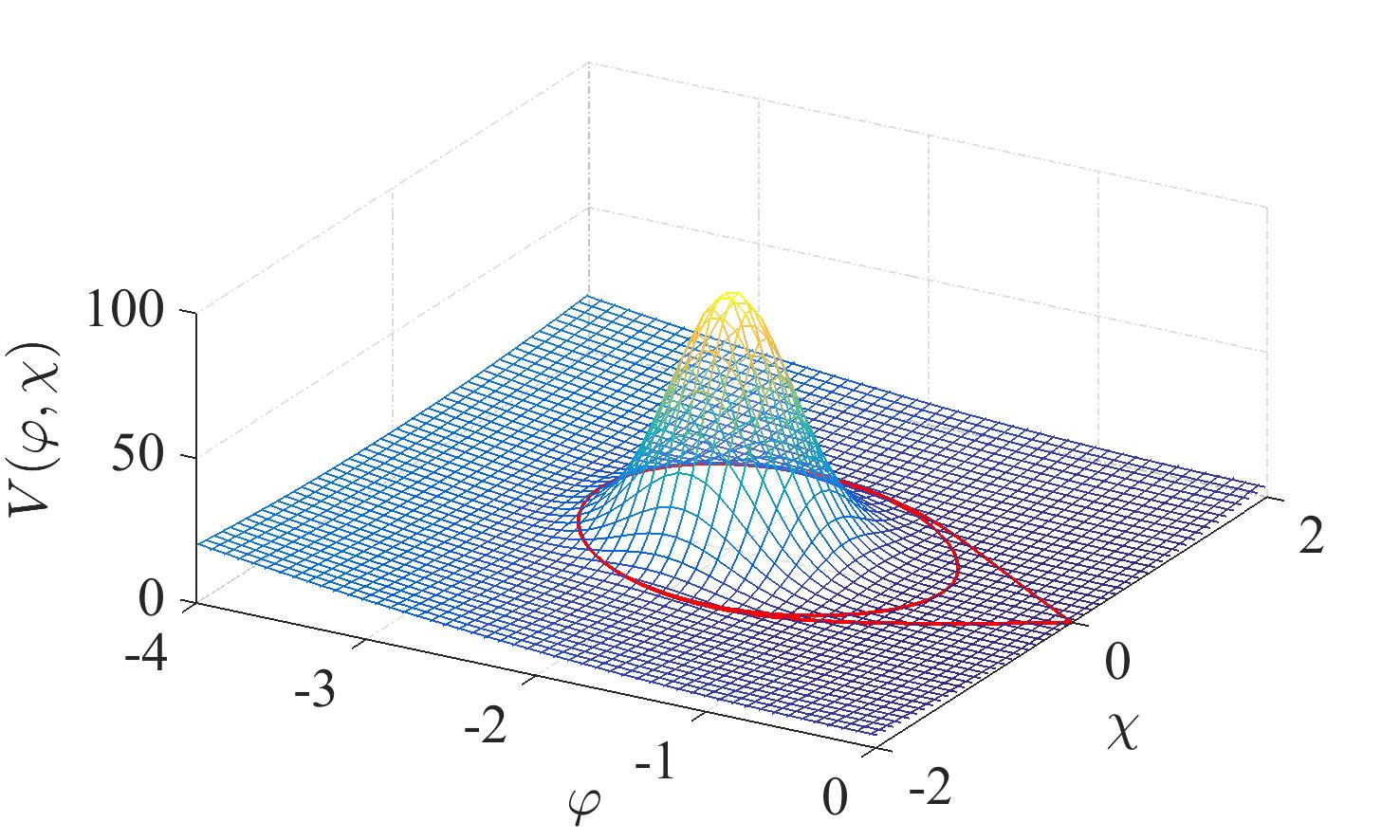}
	  \label{fig:Potential-23N}
	 }
	\caption{Steady stationary configuration of the fields $\varphi$, $\chi$ on the potential surface \eqref{eq:potential} at various winding numbers $N$.}
	\label{fig:Potential+N}
\end{center}
\end{figure}

Equations \eqref{eq:Equations}, \eqref{eq:BondCond}, \eqref{eq:InCond}, and \eqref{eq:Theta} form a well-posed problem. Equations \eqref{eq:Equations} are inhomogeneous hyperbolic equations of the second order with the periodic boundary conditions \eqref{eq:BondCond}, which can be solved by the standard methods. We employed an implicit version of the finite difference method, complemented with the modified tridiagonal matrix algorithm, see, e.g., \cite{book:Samarski}. We used the following values of the parameters of the potential \eqref{eq:potential}: $a = 50$, $b = 5$, $c = 5$, $d = 1$, $\varphi_0^{} = -2$, $\chi_0^{} = 0$ (note that the saddle point has coordinates $(-2.9,0)$). Furthermore, we set $R = \sqrt{\varphi_0^2 + \chi_0^2}$, see figure \ref{fig:Potential+InCond} (right panel). 

\begin{figure}[th!]
\begin{center}
	\subfigure[$N=1$.]{	\includegraphics[width=0.47\textwidth]{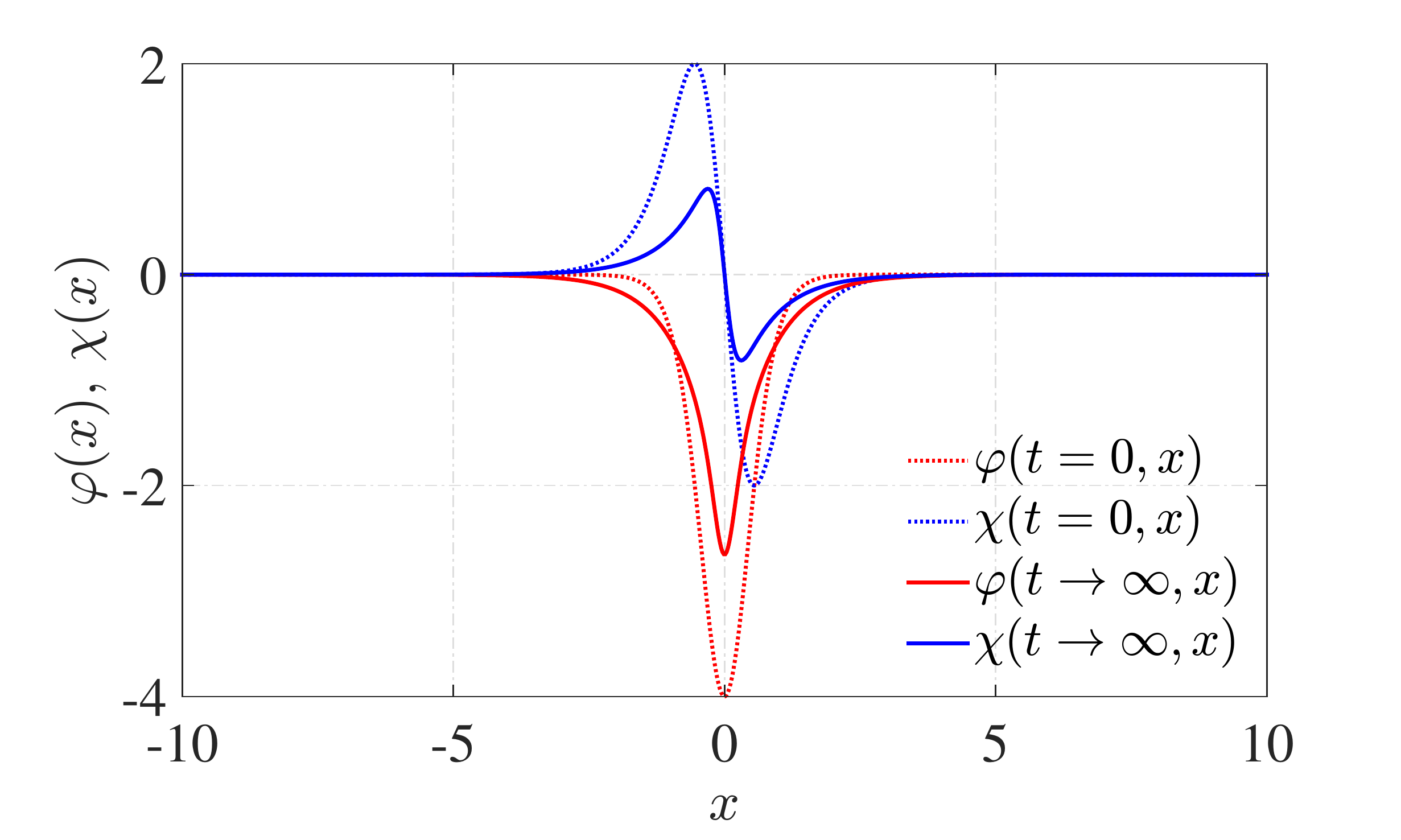}
		\label{fig:Fields-1N}
	}
	\subfigure[$N=2$. For $N>2$ the number of ``humps'' is increasing proportional to $N$ (see \cite{GKR_Sol_proc} for details).]{	\includegraphics[width=0.47\textwidth]{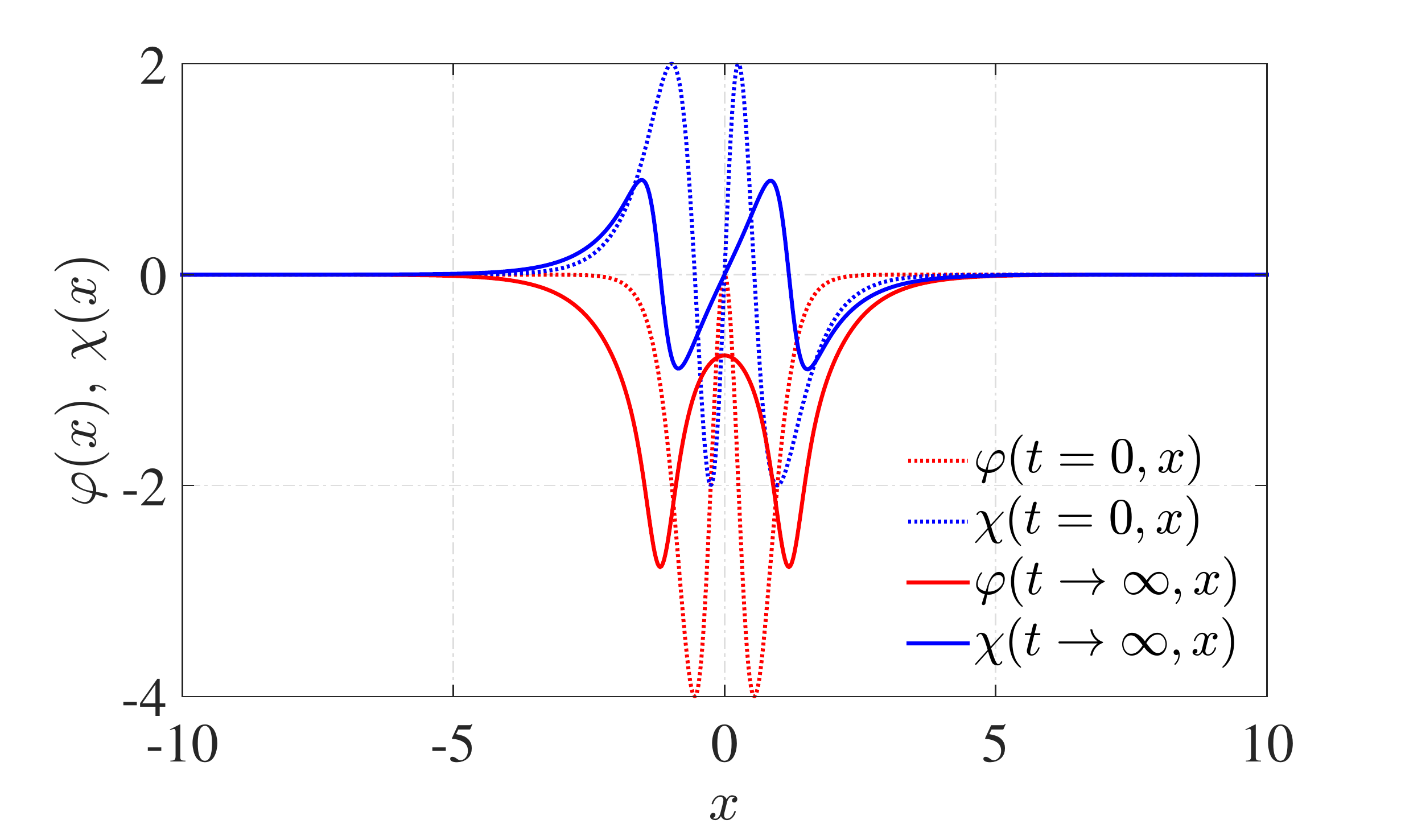}
		\label{fig:Fields-2N}
	}
\end{center}
	\caption{Spatial distributions of the fields $\varphi$, $\chi$ at the initial ($t=0$) and final ($t\to+\infty$) moments of time for the potential \eqref{eq:potential}, corresponding to various winding numbers $N$.
    }
	\label{fig:Fields+Density}
\end{figure}

We performed the numerical simulation of the evolution of the initial configuration \eqref{eq:InCond}, \eqref{eq:Theta} for $N=1,2,3,4$ and confirmed possibility of soliton production for all $N$, which was shown earlier in \cite{GKR_Sol_proc}. Our results are presented in figures \ref{fig:Potential+N} and \ref{fig:Fields+Density}. At all considered winding numbers we observed the tightening of the loop around the maximum of the potential. This means that a space trajectory connecting a point of the domain $\mathcal B$ with a point of $\mathcal U$ necessarily goes through a domain wall (at $N=1$) or a series of domain walls (at $N\ge 2$). 

If the height of the potential maximum is small enough, the fields overcome the maximum and tend to the absolute minimum in the whole space. We obtained that solitons with the winding number $N=2$ survive at $a\gtrsim 84$. For example, in figs.~\ref{fig:Potential-23N} and \ref{fig:Fields-2N} we used $a = 84$. On the other hand, at $a\lesssim 50$ (with the same values of the other parameters) the solitons are not formed (unwind).

\begin{figure}[t]
	\centering
	\includegraphics[width=0.6\textwidth]{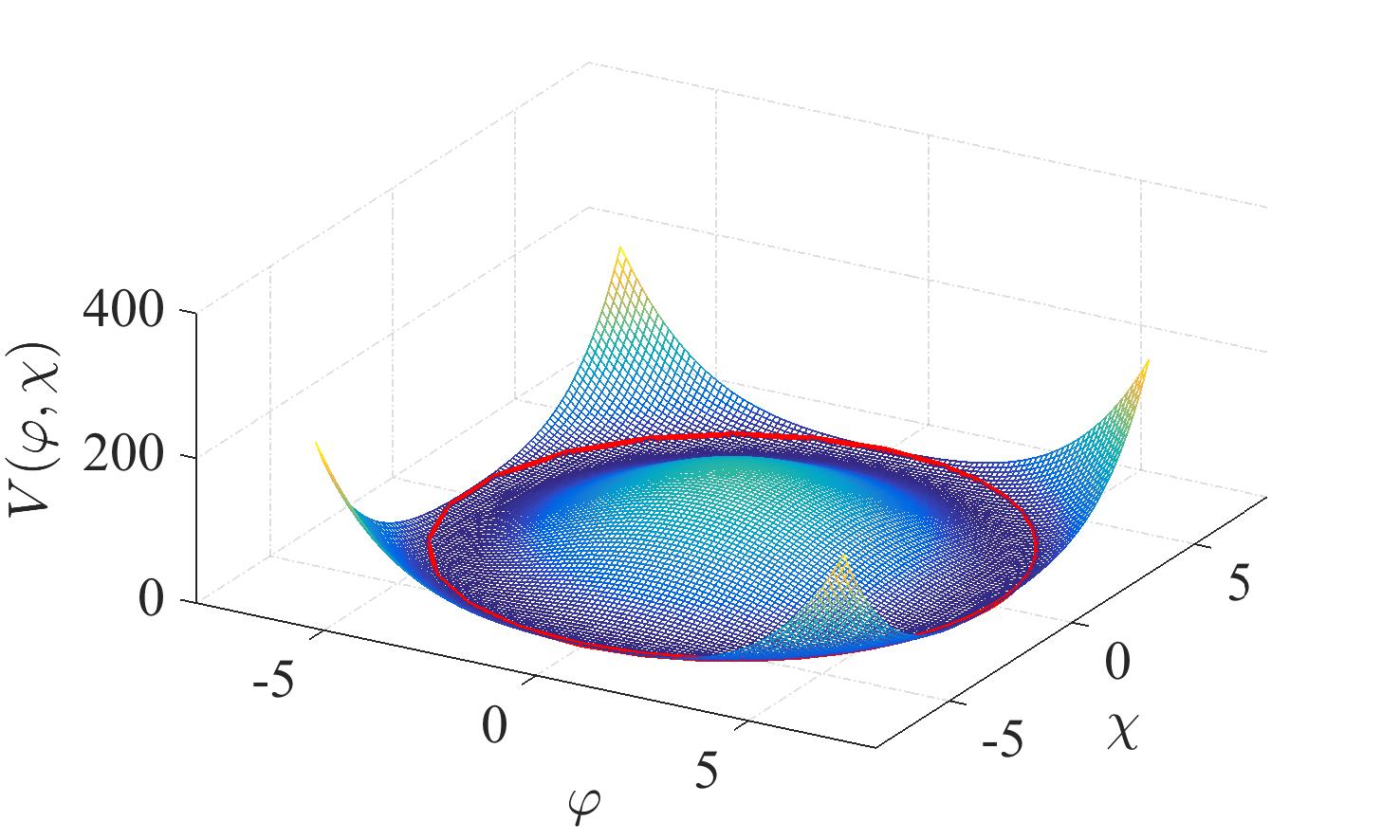}
	\caption{Initial state of the fields $\varphi$, $\chi$ on the potential surface \eqref{eq:Potential_Mex}. The parameter values are listed under \eqref{eq:Potential_Mex}.}
    \label{fig:Potential_Mex}
\end{figure}

Let us now discuss the conditions under which the change of the winding number becomes possible during the classical evolution of the fields. To facilitate the analysis, we consider the well known tilted Mexican hat potential \cite{Dolgov}. For two real scalar fields $\varphi$ and $\chi$ this potential has the form:
\begin{equation}
	V(\varphi,\chi) = \lambda \left(\varphi^2 + \chi^2 - \frac{g^2}{2}\right)^2 
		+ \Lambda^4 \left(1 - \frac{\varphi}{\sqrt{\varphi^2 + \chi^2}}\right).
	\label{eq:Potential_Mex}
\end{equation}
In our calculations below we used $g=8$ and $\Lambda=5\cdot 10^{-13}$, see figure \ref{fig:Potential_Mex}.
Solving the equations of motion \eqref{eq:Equations} with the boundary conditions \eqref{eq:BondCond} and the initial conditions \eqref{eq:InCond} with $\varphi_0^{}=\chi_0^{}=0$ and $R=(1+\varkappa)g/\sqrt{2}$ for the potential \eqref{eq:Potential_Mex} ($\varkappa=0$ corresponds to the potential minimum valley; in our calculation we used $\varkappa=0.1$ for the initial conditions), we observed  the formation of a soliton analogous to those studied above. 

\begin{figure}[h!]
	\centering
    \subfigure[$\lambda=0.1$, $\Lambda=5\cdot10^{-13}$, $g=7$.]{
	  \includegraphics[width=0.45\textwidth]{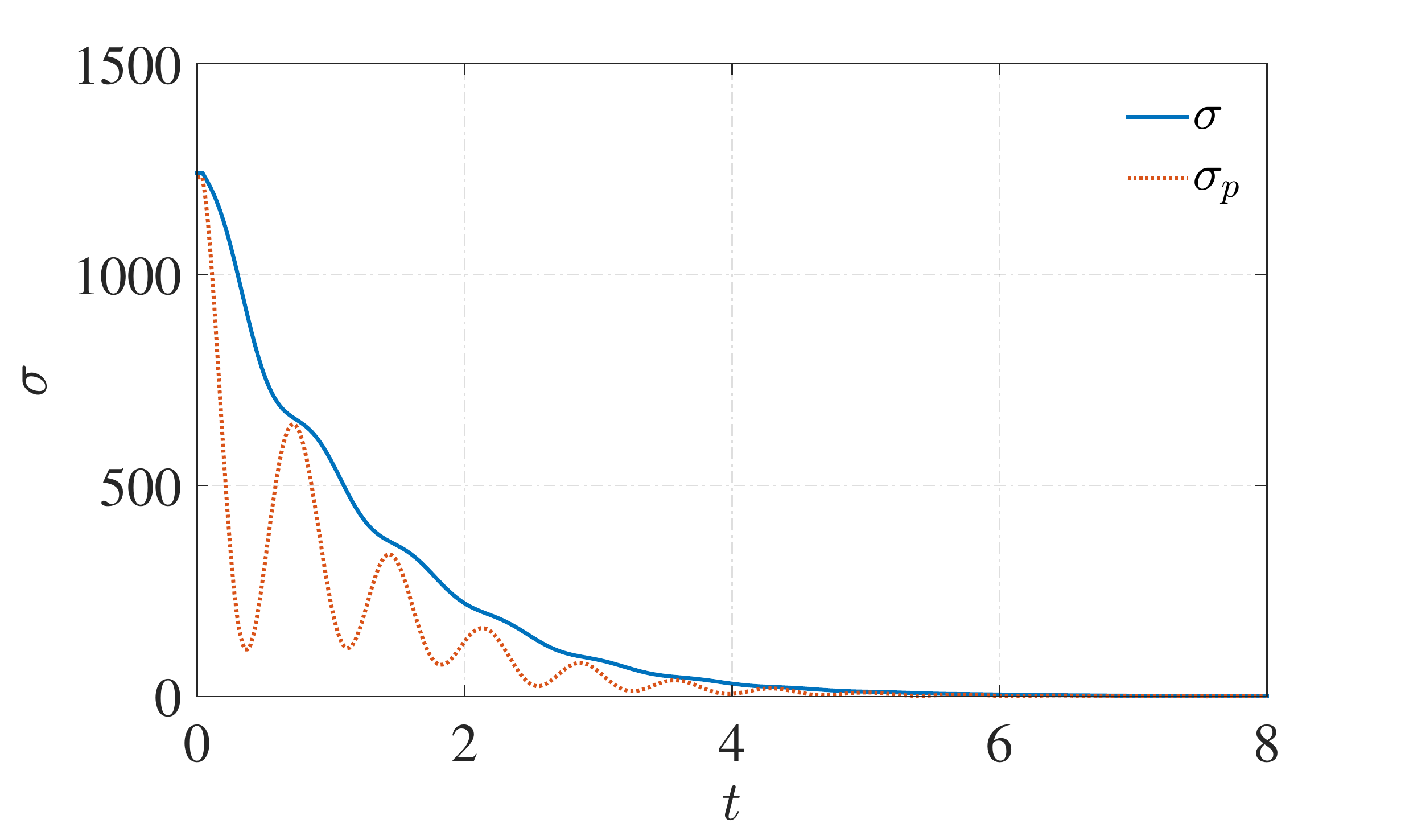}
     }
    \subfigure[$\lambda=0.1$, $\Lambda=5\cdot10^{-13}$, $g=8$.]{
	  \includegraphics[width=0.45\textwidth]{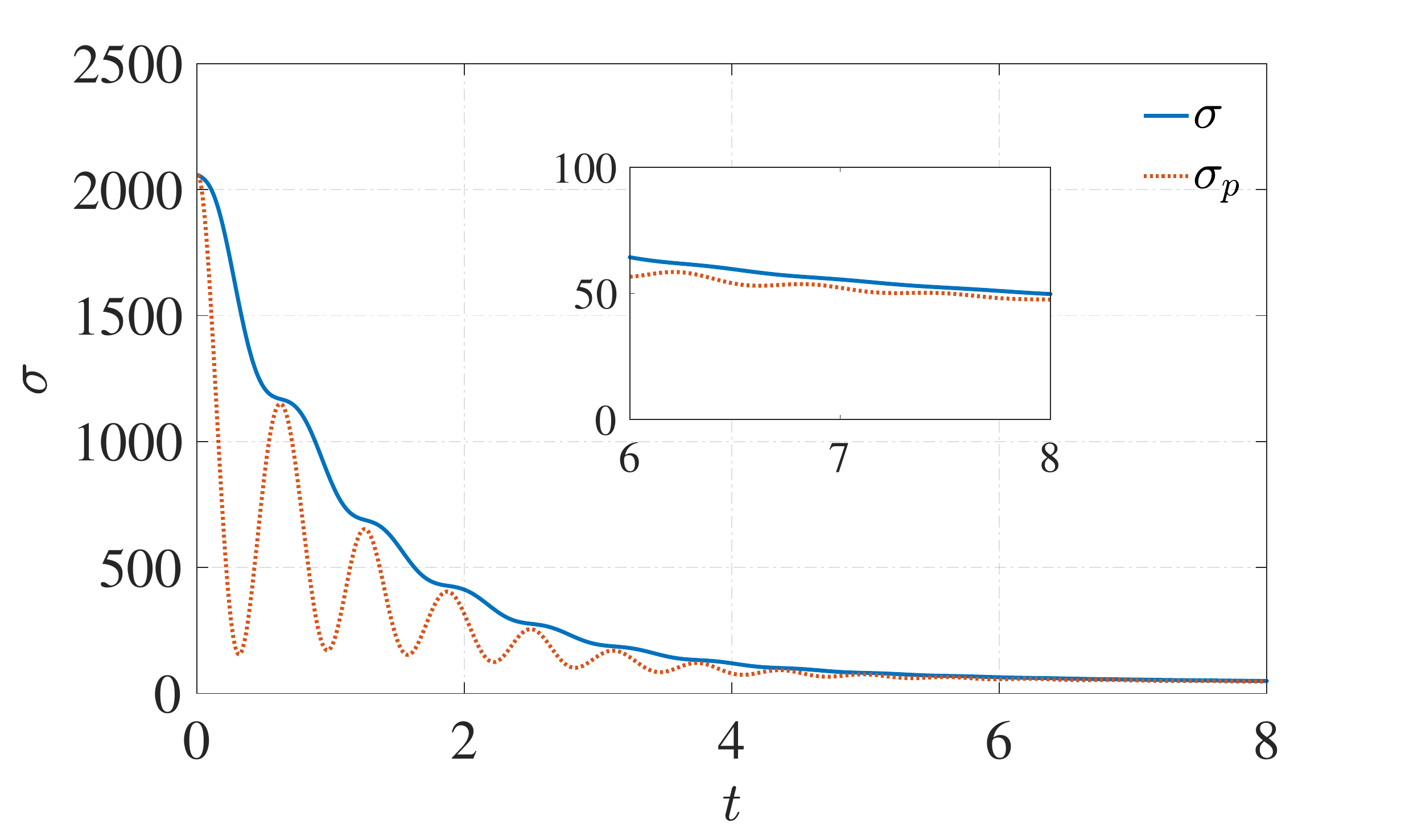}
     }
\caption{Typical time-dependence of the total surface energy density $\sigma$ (the total energy of the field configuration) and the potential surface energy density $\sigma_\mathrm{p}$ for the potential \eqref{eq:Potential_Mex} at $\lambda=0.1$, $\Lambda=5\cdot10^{-13}$, $g=7$ and $g=8$ (in $H_\mathrm{I}$ units) and the initial conditions \eqref{eq:InCond} with $R=(1+\varkappa)g/\sqrt{2}$. The parameters are chosen in such a way that the soliton doesn't overcome the maximum of the potential (i.e.\ doesn't unwind). }
	\label{fig:E_t_viol}
\end{figure}

As a check of our numerical results, we also calculated the total surface energy density $\sigma$ of the field configuration. As expected, it monotonically decreases with time, while the potential energy density oscillates, see figure \ref{fig:E_t_viol}.  If a soliton decays, the final energy tends to zero due to the friction (left panel). On the contrary, the energy tends to a constant as it should be for the soliton formation (right panel). The left and the right panels differ by slight variation of the parameter $g$.

Thus we conclude that the initial field configuration \eqref{eq:InCond} evolves differently depending on the shape of the potential. If the height of the potential maximum is small, the formation of the configurations with non-zero winding number is not observed. On the other hand, if the height of the maximum is bigger than some critical value, the configurations with non-zero winding numbers can be formed. As is discussed in the next section, the formation of the topologically non-trivial configurations could substantially affect the process of the Universe formation.

\section{Large density perturbations in the early Universe}
\label{sec:Models}

In the previous sections we have considered the potentials \eqref{eq:potential} and \eqref{eq:Potential_Mex}. We have demonstrated that depending on the height of the potential maximum closed walls could be formed just after the end of the inflation. In this section we discuss influence of this result on the processes in the early Universe.

Let us show that the relative density excess $\delta\rho/\rho$ can be of order of unity even for large scales. In \cite{Rub_GalNucl} we considered mechanism of formation of closed walls based on the quantum fluctuations during inflation around a saddle point. As it was shown in \cite{KhlopRubSakh}, these fluctuations lead to the fractal-like structure of walls. As a result, the surface area and consequently its mass are increased by a factor $\zeta^N$ where $N$ is the number of e-folds and $\zeta\sim 1\div 2$. As it was shown in \cite{Widrow}, the crinkles could live long enough to prevent BH formation as it would be in the case of spherical wall and the Minkowski background. In more realistic case of FRW universe, the wall crinkles transfer their energy into surrounding media. The produced excess of the energy density may be considered as non-gaussian fluctuations that influence the CMB spectrum and the galaxy formation. Let us make some estimations.

The  wall energy is approximately $E_\mathrm{w}\sim 4\pi\sigma R_\mathrm{h}^2\zeta^N$. Here $R_\mathrm{h}$ is the scale of the closed domain wall at the moment $t_h$ of the horizon crossing. Hence, $\delta\rho\sim E_\mathrm{w}/\left(\displaystyle\frac{4\pi}{3}R_\mathrm{h}^3\right)$. If the wall was formed at the e-fold number $N$ then it will be crossed by horizon having the size  $R_\mathrm{h}\sim H^{-1}e^{2N}$ (at the radiation dominated stage). The matter density is $\rho\sim T^4 \sim M_\mathrm{Pl}^2/R_\mathrm{h}^2$. Gathering everything, we obtain:
\begin{equation}
 \label{formula}
 \frac{\delta\rho}{\rho}\sim \frac{\sigma}{M_\mathrm{Pl}^2 H}\,\zeta^N e^{2N}.
\end{equation}
After the wall crosses the horizon it starts shrinking that leads with necessity to strong inhomogeneities in the density distribution. For example, suppose that $\zeta=1.2$. Then, for condition $\displaystyle\frac{\delta\rho}{\rho} \simeq 1$ formula \eqref{formula} gives $N\simeq 37$ for $\sigma=4\Lambda^2 g\simeq 10^{16}$~GeV$^3$ at the parameter values $\Lambda$ and $g$ used in the previous section. This scale is much less than the galaxy scale. 

The result is very sensitive to the parameter $\zeta$, which is poorly known. For example, consider the surface energy density $\sigma=10^4$~GeV$^3$. The energy density excess \eqref{formula} at the galaxy scale ($N\sim45$) is negligibly small --- $\displaystyle\frac{\delta\rho}{\rho}\sim 10^{-8}$ for $\zeta=1$~--- and of the order of 1, $\displaystyle\frac{\delta\rho}{\rho}\sim 1$, for $\zeta=1.5$. 

The energy of crinkles starts to flow to the surrounding media when the horizon reaches the scale of the crinkle. The latter is much smaller than the scale of the domain wall. Hence the crinkle could disappear before the wall starts to evolve as a whole. The estimations performed in \cite{vilenkin01,Kibble} indicate that perturbations of the scales smaller than $\sqrt{G\sigma}t_\mathrm{h}^{3/2}$ are damped to the moment $t_\mathrm{h}$ of the horizon crossing the scale of the wall. The energy flow from the crinkles to the ambient media leads to its heating. In turn, the temperature growth makes the energy flow more intensive. The impact of this effect on the mass of the black hole and the very fact of its formation are strongly model dependent. First estimates indicate that the effect is not very prominent. For example, suppose that the average energy density excess of a wall is $\delta\rho \sim \rho$, see the previous paragraph. In this case, the ambient media would be heated in $2^{1/4}$ times and this effect can be neglected.

Therefore the presence of saddle point(s) leads to heated overdensed areas of large scale. Notice that the role of the crinkles appears to be small if the walls are the result of the quantum tunneling \cite{DengVil2017}. The physical explanation is that the fluctuations of their form are suppressed by the exponential factor $e^{-S_\mathrm{E}}$ where $S_\mathrm{E}$ is the Euclidean action. 

Another interesting possibility has been discussed since 1984 \cite{Ipser}, where it was revealed that closed domain walls could form black holes after shrinking. At the first glance, a BH mass should be approximately equal to the mass of the closed wall. In reality, there are several effects that strongly influence the result \cite{DengVil2017, Garriga2016}. A wall for which $t_\mathrm{h} < t_{\sigma}\equiv \displaystyle\frac{M_\mathrm{Pl}^2}{2\pi\sigma}$ is called “subcritical”, and its gravitational field can be safely neglected before horizon crossing. If its form is near spherical, it shrinks to a size smaller than the corresponding Schwarzschild radius, forming a black hole with ordinary internal structure. If the surface energy density $\sigma$ is large so that $t_\mathrm{h} > t_{\sigma}$, the black holes are also being formed for an exterior observer. Meanwhile their interior inflates, forming baby universes. The latter are connected to the exterior FRW Universe by wormholes \cite{Garriga2015}. The final masses of the black holes depending on the initial data and parameters of the model are analyzed in \cite{Garriga2015, Garriga2016, DengVil2017}.

Let us address the question, whether the massive PBHs could be formed in spite of the wall disappearance after the end of the inflation. To clarify the subject, we perform some estimations and find conditions that lead to the BHs formation within this scenario. The gravitational radius of the domain wall is of the order of $r_\mathrm{g} =2GE_\mathrm{w}\simeq 8\pi\sigma l_\mathrm{h}^2/M_\mathrm{Pl}^2$ for spherical wall ($\zeta =1$), where the horizon size $l_\mathrm{h}$ ($\simeq  R_\mathrm{h}$) crosses the domain wall size at the moment $t\sim l_\mathrm{h}$. The wall energy $E_\mathrm{w}$ is emitted in the form of waves of the scalar fields if the fields overcome the potential maximum and fluctuates near the potential minimum. The waves are emitted both outside and inside the domain. The latter are concentrated on the scale of the order of the wavelength $l_\mathrm{w}$. Hence the condition for the BH formation is
\begin{equation}\label{ineq}
l_\mathrm{w} < r_\mathrm{g} \simeq \frac{8\pi\sigma l_\mathrm{h}^2}{M_\mathrm{Pl}^2}.
\end{equation}
This inequality may be used to estimate the horizon size $l_\mathrm{h}$. Let us return to the potential \eqref{eq:Potential_Mex} with the values of the parameters used in figure \ref{fig:E_t_viol}. Assume that if the fields overcome the potential maximum then they start to oscillate around the minimum $r=\sqrt{\varphi^2 + \chi^2}=0$ in the radial direction. The frequency of such oscillations is of the order of $1/m_\mathrm{r}$ where $m_\mathrm{r}=2\sqrt{\lambda} g$ is the mass of the radial oscillations. Hence the wavelength emitted during the oscillations is of the order of $l_\mathrm{w}\sim 1/m_\mathrm{r}=(2\sqrt{\lambda}\, g)^{-1}$. The surface energy density of the domain wall is $\sigma =4\Lambda^2 g$. We can estimate the horizon scale using \eqref{ineq}:
\begin{equation}
l_\mathrm{h} \gtrsim \frac{M_\text{Pl}}{\lambda^{1/4}\Lambda g}\sim 10^4 \text{ GeV}^{-1},
\end{equation}
that is not very serious restriction.
Therefore, the PBHs are intensively produced even if the domain walls finally disappear emitting radiation according to the second scenario. Accurate calculation of the PBH mass spectra for disappeared closed walls will be the subject of further research.

\section{Conclusion}
\label{sec:Conclusion}

We have considered the evolution of the non-trivial field configurations after the end of the inflation in the model with the potential having saddle point(s). We have shown that there are substantially different scenarios of the field evolution.

The first scenario is widely discussed in the framework of the random Gaussian landscape \cite{Vil} and multi-stream inflation \cite{li01}. It was shown that the field dynamics during the inflation leads to inhomogeneities in the CMB radiation temperature fluctuations that could be observed in future. As we noticed in the Introduction, models with many peak potentials could contain saddle points that lead to serious consequences discussed above. 

The second scenario has been elaborated in this paper. We have shown that the potentials with a saddle point could lead to the closed wall formation even if there is only one potential minimum.

Realization of the first or the second scenario depends on the parameters of the potential and on the initial conditions. If the second scenario takes place, strong inhomogeneities in the CMB spectrum and large energy density fluctuations appear in the early Universe. Possible formation of the massive PBHs is also discussed in the paper. Our estimations show that PBHs can be formed even if the domain walls disappear (the second scenario). This result worths to be thoroughly investigated in future. For example, it would be interesting to check if the PBHs can also be formed in the multi-stream inflation.

The set of the field configurations that could evolve into BHs can be split into disjoint equivalence classes that are identified by the winding number of the configuration. We have provided simple formula for analytical and numerical calculation of the winding number of arbitrary field configuration. If the height of the potential is not very large, the winding number could change during the field evolution.

The new type of solitons studied here may be also useful as a mechanism of production of black holes with complex mass spectra \cite{Konopl,RubKhlopSakh_PBHProd, KhlopRubSakh,Khlop,Belotsky,PBH,PBH-Kuhnel} and could be used as a part of the solution of a lot of cosmological problems: high-$z$ quasars and supermassive black holes within galactic nuclei \cite{Konopl,RubKhlopSakh_PBHProd,Rub_GalNucl,KhlopRubSakh,Dokuch_Quasars,Khlop,Dokuch,PBH,Grob1}, dark matter \cite{Belotsky,Grob2,PBH,PBH-Carr1,PBH-Dolgov,Gani.IJMPD.2015,Gani.PP.2015}, as well as astrophysical problems: unidentified sources of gamma-radiation \cite{Belotsky_PBH_gamma_1,Belotsky_PBH_gamma_2,Belotsky}, reionization \cite{Belotsky,PBH-reion1,PBH-reion2,PBH-reion3}, the deficiency of intermediate-mass black holes \cite{Dokuch2,Grob1}.

The fluctuating mechanism discussed in this paper leads to the PBH formation due to the closed walls collapse. The PBH mass spectrum is the result of complex processes that take place just after the inflation is finished and is strongly model dependent. Indeed, most of the walls are created having the form far from the spherical one. Their small inhomogeneities are damped transferring the energy to the surrounding media and heating it. The larger perturbations of the wall surface survive to the moment of the horizon crossing. The supercritical walls ($t_\mathrm{h} \gg t_{\sigma}$) form baby universes with wormholes. At the same time, the nonspherical collapse with the BHs in the final stage has been discussed in \cite{Bernal,Bantilan}. Moreover, as was shown in \cite{RubKhlopSakh_PBHProd} black holes are created together with substantial amount of BHs of smaller masses. The consequent evolution of such PBH cluster depends on its space/mass distribution, parameters of the Lagrangian, and initial conditions.

In conclusion, we emphasize that our consideration is limited to two-dimensional domain walls in the three-dimensional space, whereas the formation and evolution of, e.g., string-type configurations would be the subject of future study. 

\section*{Acknowledgments}

The authors would like to thank K.~Belotsky for useful discussions. The work fulfilled in the framework of MEPhI Academic Excellence Project (contract \textnumero~02.a03.21.0005, 27.08.2013) and according to the Russian Government Program of Competitive Growth of Kazan Federal University. The work of S.G.R. was also supported by the Ministry of Education and Science of the Russian Federation, Project \textnumero~3.4970.2017/BY.

\end{document}